\documentclass{iau}
\usepackage{graphicx}

\title[Large central mass-to-light ratios in massive early-type
 galaxies]{Further evidence for\\ large central mass-to-light ratios in\\
 massive early-type galaxies}

\author[E. M. Corsini et al.]{E. M. Corsini$^{1,2}$, G. A. Wegner$^3$,
  J. Thomas$^4$, R. P. Saglia$^4$,\\ R. Bender$^{4,5}$ and
  S. B. Pu$^6$}

\affiliation{$^1$Dipartimento di Fisica e Astronomia, Universit\`a di Padova, 
   Padova, Italy\\ email: {\tt enricomaria.corsini@unipd.it} \\[\affilskip]
$^2$INAF--Osservatorio Astronomico di Padova, Padova, Italy \\[\affilskip]
$^3$Department of Physics and Astronomy, Dartmouth College, Hanover, 
   NH, USA \\[\affilskip]
$^4$Max-Planck-Institut f\"ur extraterrestrische Physik,  
   Garching, Germany \\[\affilskip]
$^5$Universit\"ats-Sternwarte M\"unchen, 
   M\"unchen, Germany \\[\affilskip]
$^6$The Beijing No. 12 High School, Beijing, China}

\pubyear{2013}
\volume{295}  
\pagerange{xx--xx}
\jname{The intriguing life of massive galaxies}
\editors{D. Thomas, A. Pasquali \& I. Ferreras, eds.}
\begin{document}

\maketitle

\begin{abstract}
We studied the stellar populations, distribution of dark matter, and
dynamical structure of a sample of 25 early-type galaxies in the Coma and
Abell~262 clusters.
We derived dynamical mass-to-light ratios and dark matter densities
from orbit-based dynamical models, complemented by the ages,
metallicities, and $\alpha$-elements abundances of the galaxies from
single stellar population models.
Most of the galaxies have a significant detection of dark matter and
their halos are about 10 times denser than in spirals of the same
stellar mass. Calibrating dark matter densities to cosmological
simulations we find assembly redshifts $z_\mathrm{DM}\approx 1-3$.  
The dynamical mass that follows the light is larger than expected for
a Kroupa stellar initial mass function, especially in galaxies with
high velocity dispersion $\sigma_\mathrm{eff}$ inside the effective
radius $r_\mathrm{eff}$.
We now have 5 of 25 galaxies where mass follows light to
$1-3\,r_\mathrm{eff}$, the dynamical mass-to-light ratio of all the
mass that follows the light is large ($\approx\,8-10$ in the
Kron-Cousins $R$ band), the dark matter fraction is negligible to
$1-3\,r_\mathrm{eff}$.
This could indicate a `massive' initial mass function in massive
early-type galaxies. Alternatively, some of the dark matter in massive
galaxies could follow the light very closely suggesting a significant
degeneracy between luminous and dark matter.
\end{abstract}

\firstsection 

\section{Introduction}

In the past years we studied the stellar populations, mass
distribution, and orbital structure of a sample of early-type galaxies
in the Coma cluster with the aim of constraining the epoch and
mechanism of their assembly.

The surface-brightness distribution was obtained from ground-based and
HST data. The stellar rotation, velocity dispersion, and the $H_3$ and
$H_4$ coefficients of the line-of-sight velocity distribution were
measured along the major axis, minor axis, and an intermediate
axis. In addition, the line index profiles of Mg, Fe and H$\beta$ were
derived (\cite[Mehlert \etal\ 2000; Wegner \etal\ 2002; Corsini
  \etal\ 2008]{Mehlert2000, Wegner2002, Corsini2008}). Axisymmetric
orbit-based dynamical models were used to derive the mass-to-light
ratio $\Upsilon_\ast$ of all the mass that follows the light and the
dark matter (DM) halo parameters in 17 galaxies (\cite[Thomas
  \etal\ 2005, 2007a,b, 2009a,b]{Thomas2005, Thomas2007a, Thomas2007b,
  Thomas2009a, Thomas2009b}). The comparison with masses derived
through strong gravitational lensing for early-type galaxies with
similar velocity dispersion and the analysis of the ionized-gas
kinematics gave valuable consistency checks for the total mass
distribution predicted by dynamical modeling (\cite[Thomas
  \etal\ 2011]{Thomas2011}). The line-strength indices were analyzed
by single stellar-population models to derive the age, metallicity,
$\alpha$-elements abundance, and mass-to-light ratio
$\Upsilon_\mathrm{Kroupa}$ (or $\Upsilon_\mathrm{Salpeter}$ depending
on the adopted initial mass function, IMF) of the galaxies
(\cite[Mehlert \etal\ 2003]{Mehlert2003}).

More recently, we have performed the same dynamical analysis for 8
early-type galaxies of the nearby cluster Abell~262 (\cite[Wegner
  \etal\ 2012]{Wegner2012}). The latter is far less densely populated
than Coma cluster and it is comparable to the Virgo cluster. Moreover,
while the Coma galaxies were selected to be mostly flattened, the
Abell~262 galaxies we measured appear predominantly round on the sky.

\section{Results}

\subsection{Evidence for halo mass not associated to the light} 

In the Coma galaxy sample the statistical significance for DM halos is
over $95\%$ for 8 (out of 17) galaxies (\cite[Thomas
  \etal\ 2007b]{Thomas2007b}), whereas the Abell~262 sample reveals 4
(out of 8) galaxies of this kind (\cite[Wegner
  \etal\ 2012]{Wegner2012}). In Coma, we found only one galaxy
(GMP~1990) with $f_\mathrm{halo}\,\approx\,0$, i.e., with a negligible
halo-mass fraction of the total mass inside $r_\mathrm{eff}$. This is
also the case of 4 galaxies in Abell~262 (NGC~703, NGC~708, NGC~712,
and UGC~1308).
The evidence for a DM component in addition mass that follows light is
not directly connected to the spatial extent of the kinematic data,
degree of rotation, or flattening of the system.  There is no
relationship with the age, metallicity, and $\alpha$-elements
abundance of the stellar populations.

We can not discriminate between cuspy and logarithmic halos based on
the quality of the kinematic fits, except for NGC~703 where the
logarithmic halo fits better. Still the majority of cluster early-type
galaxies have $2-10$ times denser halos than local spirals (\cite[e.g.,
  Persic \etal\ 1996]{Persic1996}), implying a $1.3-2.2$ times higher
$(1+z_\mathrm{DM})$ assuming
$\langle\,\rho_\mathrm{DM}\,\rangle\,\sim\,(1+z_\mathrm{DM})^3$, where
$z_\mathrm{DM}$ is the formation redshift of the DM halos. Thus, if
spirals typically formed at $z_\mathrm{DM}\,\approx\,1$, then cluster
early-type galaxies assembled at $z_\mathrm{DM}\,\approx\,1.6-3.4$.

Averaging over all galaxies, we find that a fraction of $\langle
f_\mathrm{halo} \rangle\,=\,0.2$ of the total mass inside
$r_\mathrm{eff}$ is in a DM halo distinct from the light.  Similar
fractions come from other dynamical studies employing spherical models
(\cite[e.g., Gerhard \etal\ 2001]{Gerhard2001}). The Coma and
Abell~262 galaxies show an anti-correlation between
$\Upsilon_\ast/\Upsilon_\mathrm{Kroupa}$, i.e. the ratio between the
dynamical and stellar population mass-to-light ratios, and
$f_\mathrm{halo}$ (Fig. \ref{fig:mlplot}, left panel). Galaxies where
the dynamical mass following the light exceeds the Kroupa value by far
($\Upsilon_\ast/\Upsilon_\mathrm{Kroupa}\,>\,3$) seem to lack matter
following the halo distribution inside $r_\mathrm{eff}$ ($\langle
f_\mathrm{halo} \rangle\,\approx\,0$). Not so in galaxies near the
Kroupa limit ($\Upsilon_\ast/\Upsilon_\mathrm{Kroupa}\,<\,1.4$), where
the dark-halo mass fraction is at its maximum ($\langle
f_\mathrm{halo} \rangle\,=\,0.3$).

\subsection{Mass that follows the light}

\begin{figure}[t]
\begin{center}
\includegraphics[width=5cm]{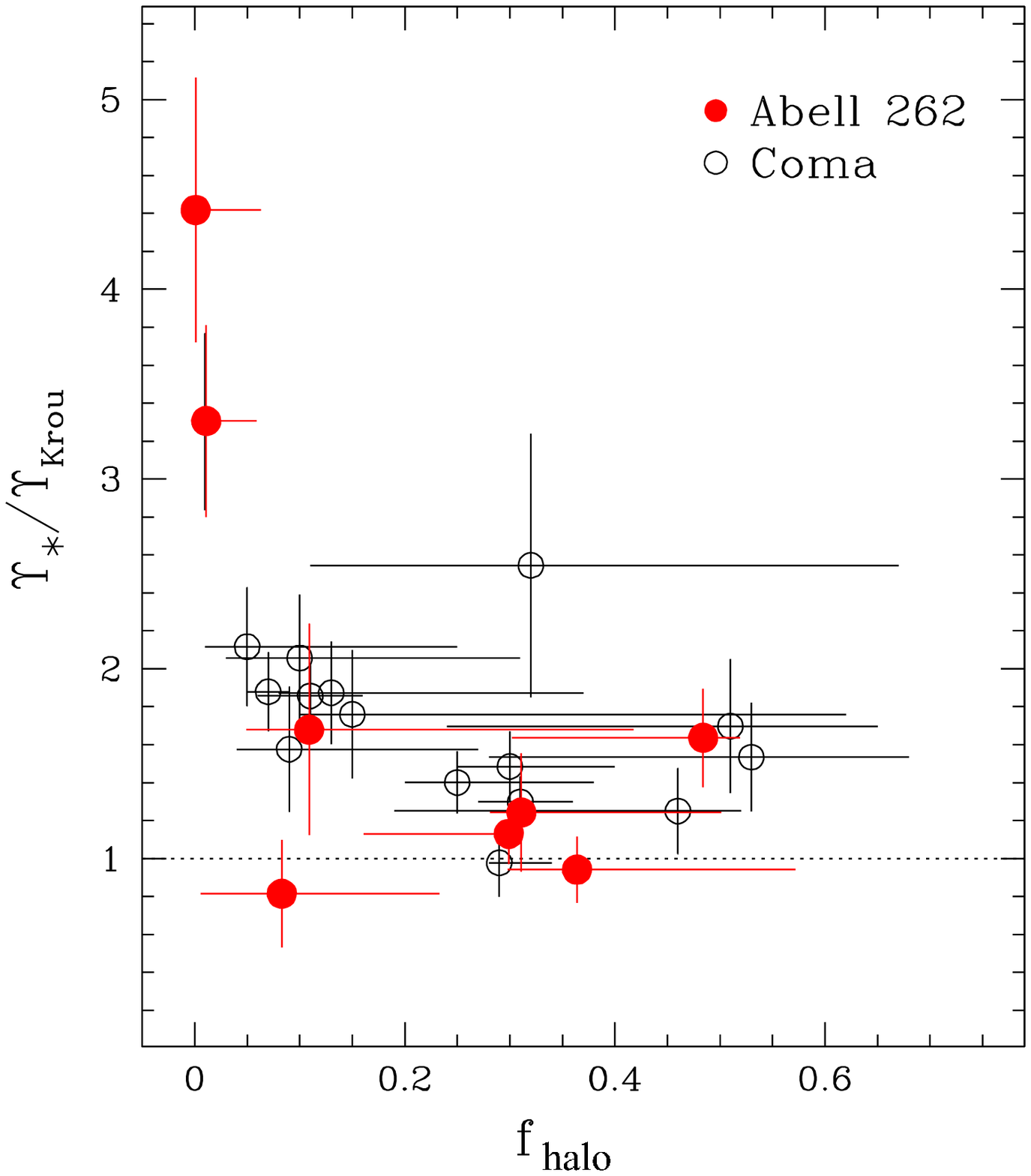}
\includegraphics[width=5cm]{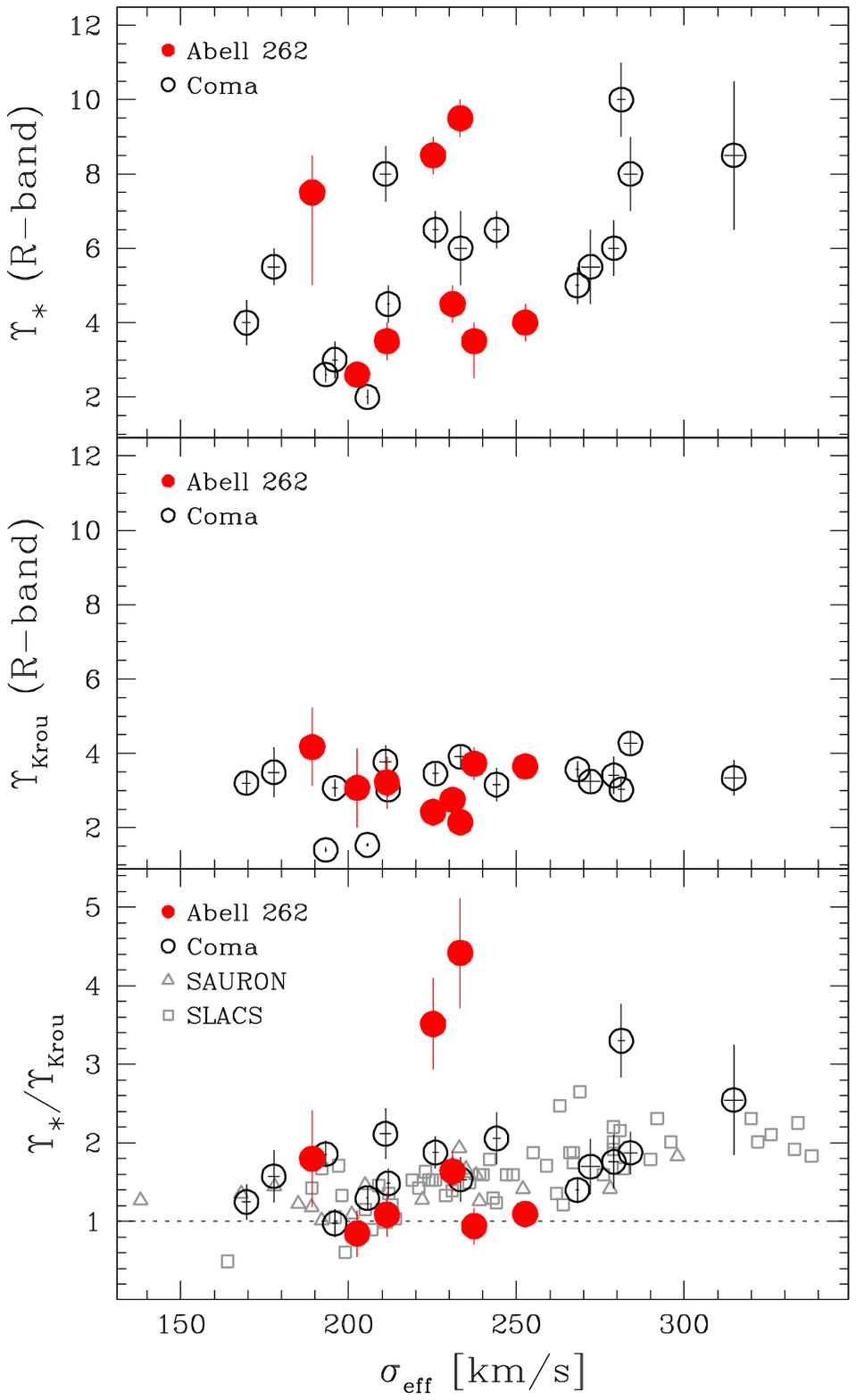} 
\caption{Left: Ratio of dynamical $\Upsilon_\ast$ to
  stellar-population $\Upsilon_\mathrm{Kroupa}$ as a function of
  $f_\mathrm{halo}$, i.e., the halo-mass fraction of the total mass
  inside $r_\mathrm{eff}$, for galaxies in Coma (\cite[open circles,
    Thomas \etal\ 2011]{Thomas2011}) and Abell~262 (\cite[filled
    circles, Wegner \etal\ 2012]{Wegner2012}).
  $r_\mathrm{eff}$. $\Upsilon_\ast/\Upsilon_\mathrm{Kroupa}=1.6$
  corresponds to a Salpeter IMF. Right: Dynamical $\Upsilon_\ast$
  (upper panel), stellar-population $\Upsilon_\mathrm{Kroupa}$ (middle
  panel), and their ratio (bottom panel) as a function of the
  effective velocity dispersion, $\sigma_\mathrm{eff}$. In the bottom
  panel, Coma and Abell~262 galaxies are compared to SLACS
  (\cite[open squares, Treu \etal\ 2010]{Treu2010}) and SAURON
  galaxies (\cite[open triangles, Cappellari
    \etal\ 2006]{Cappellari2006}).}
  \label{fig:mlplot}
\end{center}
\end{figure}

As far as the mass-to-light ratios are concerned, the galaxies of Coma
and Abell~262 follow a similar trend.  While the dynamically
determined $\Upsilon_\ast$ increases strongly with
$\sigma_\mathrm{eff}$, i.e., the velocity dispersion averaged within
$r_\mathrm{eff}$ (Fig.~\ref{fig:mlplot}, right top panel), the stellar
population models indicate almost constant $\Upsilon_\mathrm{Kroupa}$
(Fig.~\ref{fig:mlplot}, right middle panel). This implies that the ratio
$\Upsilon_\ast/\Upsilon_\mathrm{Kroupa}$ increases with
$\sigma_\mathrm{eff}$ (Fig.~\ref{fig:mlplot}, right bottom panel).
Around $\sigma_\mathrm{eff}\,\approx\,200$ km~s$^{-1}$ the
distribution of $\Upsilon_\ast/\Upsilon_\mathrm{Kroupa}$ has a sharp
cutoff with almost no galaxy below
$\Upsilon_\ast/\Upsilon_\mathrm{Kroupa}\,=\,1$. For
$\sigma_\mathrm{eff}\,\gtrsim\,250$ km~s$^{-1}$ the lower bound of
$\Upsilon_\ast/\Upsilon_\mathrm{Kroupa}$ increases to
$\Upsilon_\ast/\Upsilon_\mathrm{Kroupa}\,\gtrsim\,2$ at $\sigma_{\rm
  eff}\,\approx\,300$ km~s$^{-1}$. Similar trends are also observed in
the SAURON sample with dynamical models lacking a separate DM halo
(\cite[Cappellari \etal\ 2006]{Cappellari2006}), in SLACS galaxies
with combined dynamical and lensing analysis (\cite[Treu
  \etal\ 2010]{Treu2010}) and, recently, in the ATLAS3d survey with
dynamical models including a DM halo (\cite[Cappellari
  \etal\ 2012]{Cappellari2012}).

\section{Discussion}

Fig.~\ref{fig:mlplot} provides strong evidence for large central
$\Upsilon_\ast$ in massive early-type galaxies. However, in all
gravity-based methods there is a fundamental degeneracy concerning the
interpretation of mass-to-light ratios. Such methods can not uniquely
discriminate between luminous and dark matter once they follow similar
radial distributions. The distinction is always based on the
assumption that the mass density profile of the DM differs from that
of the luminous matter.

One extreme point of view is the assumption that the stellar masses in
early-type galaxies are maximal and correspond to
$\Upsilon_\ast$. The immediate consequence is that the stellar IMF in
early-type galaxies is not universal, varying from Kroupa-like at low
velocity dispersions to Salpeter (or steeper) in the most massive
galaxies (\cite[see Auger \etal\ 2010, Thomas \etal\ 2011, and
  Cappellari \etal\ 2012, for a detailed discussion]{Auger2010,
  Thomas2011, Cappellari2012}). Recent attempts to measure the
stellar IMF directly from near-infrared observations point in the same
direction (\cite[see Conroy \& van Dokkum 2012, and references
  therein]{Conroy2012}).
However, we also find the galaxies with the largest
$\Upsilon_\ast/\Upsilon_\mathrm{Kroupa}$ have the lowest halo-mass
fractions inside $r_\mathrm{eff}$ and vice versa. A possible
explanation for this finding is a DM distribution that follows the
light very closely in massive galaxies and contaminates the measured
$\Upsilon_\ast$, while it is more distinct from the light in
lower-mass systems. This has been suggested elsewhere as a signature
of violent relaxation.

One option to further constrain the mass-decomposition of
gravity-based models is to incorporate predictions from cosmological
simulations that confine the maximum amount of DM that can be
plausibly attached to a galaxy of a given stellar mass.  Since
adiabatic contraction increases the amount of DM in the galaxy center,
it could be in principle a viable mechanism to lower the required
stellar masses towards a Kroupa IMF (\cite[Napolitano \etal\ 2010, but
  see also Cappellari \etal\ 2012]{Napolitano2010,
  Cappellari2012}). An immediate consequence is that some of the mass
that follows the light is actually DM, increasing the DM fraction to
about $50\%$ of the total mass inside $r_\mathrm{eff}$.

Since the (decontracted) average halo density scales with the mean
density of the universe at the assembly epoch, we derived the
dark-halo assembly redshift $z_\mathrm{DM}$ for Coma (\cite[Thomas et
  al. 2011]{Thomas2011}) and Abell~262 galaxies (\cite[Wegner et
  al. 2012]{Wegner2012}). We compared the values of $z_\mathrm{DM}$ to
the star-formation redshifts $z_\ast$ calculated from the
stellar-population ages. For the majority of galaxies
$z_\mathrm{DM}\,\approx\,z_\ast$ and their assembly seems to have
stopped before $z_\mathrm{DM}\,\approx\,1$. The stars of some galaxies
appear to be younger than the halo, which indicates a secondary
star-formation episode after the main halo assembly. The photometric
and kinematic properties of the remaining galaxies suggest they are
the remnants of gas-poor binary mergers and their progenitors formed
close to the $z_\mathrm{DM}\,=\,z_\ast$ relation.
Without trying to overinterpret the result given the assumptions, it
seems that Kroupa IMF allows us to explain the formation redshifts of
our galaxies. In addition, galaxies in Coma and Abell~262 where the
dynamical mass that follows the light is in excess of a Kroupa stellar
population do not differ in terms of their stellar population ages,
metallicities and $\alpha$-elements abundances from galaxies where
this is not the case.

Taken at face value, our dynamical mass models are therefore as
consistent with a universal IMF, as they are with a variable IMF.  If
the IMF indeed varies from galaxy to galaxy according to the average
star-formation rate (\cite[Conroy \& van Dokkum 2012]{Conroy2012})
then the assumption of a constant stellar mass-to-light ratio {\it
  inside} a galaxy should be relaxed in future dynamical and lensing
models.

\end{document}